\documentclass{article}

\usepackage[english]{babel}

\usepackage[letterpaper,top=2cm,bottom=2cm,left=3cm,right=3cm,marginparwidth=1.75cm]{geometry}

\usepackage{amsmath}
\usepackage{graphicx}
\usepackage[colorlinks=true, allcolors=blue]{hyperref}
\usepackage{booktabs}
\usepackage{tikz}
\usetikzlibrary{shapes.geometric, arrows.meta, positioning}

\title{A Process Mining-Based System For The Analysis and Prediction of Software Development Workflows}
\author{
Antía Dorado\textsuperscript{1}, Iván Folgueira\textsuperscript{1}, Sofía Martín\textsuperscript{1}, Gonzalo Martín\textsuperscript{1} ,\\ Álvaro Porto\textsuperscript{1}, Alejandro Ramos\textsuperscript{1}, John Wallace\textsuperscript{1} \\
\textsuperscript{1}Inverbis Analytics SL, Lugo, Spain
}

\begin{document}
\maketitle

\begin{abstract}
CodeSight is an end-to-end system designed to anticipate deadline compliance in software development workflows. It captures development and deployment data directly from GitHub, transforming it into process mining logs for detailed analysis. From these logs, the system generates metrics and dashboards that provide actionable insights into PR activity patterns and workflow efficiency. Building on this structured representation, CodeSight employs an LSTM model that predicts remaining PR resolution times based on sequential activity traces and static features, enabling early identification of potential deadline breaches. In tests, the system demonstrates high precision and F1 scores in predicting deadline compliance, illustrating the value of integrating process mining with machine learning for proactive software project management.
\end{abstract}

\section{Introduction}

In modern software engineering, DevOps practices have become the cornerstone of achieving continuous integration and continuous delivery (CI/CD). However, as software systems and teams scale, the complexity of development pipelines increases, making it difficult to maintain visibility over the entire workflow, from initial code contributions to deployment in production environments. While DevOps metrics, such as those popularized by the DORA framework \cite{googledorareport}, provide a valuable high-level overview of performance, they fail to capture the fine-grained process dynamics that lead to inefficiencies or bottlenecks in practice.

Process mining has emerged as a data-driven discipline capable of discovering, monitoring, and improving processes based on event logs extracted from operational systems. Although it has been widely applied in business process management and enterprise systems, its use in software development and DevOps environments remains relatively limited. However, the potential is significant: By transforming development events (commits, pull requests, builds, and deployments) into analyzable process logs, organizations can gain an increased level of transparency over how their software delivery workflows actually behave.

In this work, we present \textbf{CodeSight}, a system designed to bridge this gap by integrating process mining and predictive analytics within a real DevOps context. The platform extracts, structures, and governs data from GitHub repositories, building event logs that capture the full life cycle of software delivery, from the first commit associated with a pull request to its successful deployment in a development environment. These logs are analyzed using process mining techniques to uncover workflow patterns, identify process variants, and quantify deviations from expected paths.

Beyond descriptive analysis, CodeSight incorporates predictive modeling to estimate development times and anticipate potential delays. By experimenting with machine learning models, we demonstrate how combining process mining with predictive analytics can transform raw DevOps data into actionable insights that improve software delivery performance and foster a data-driven development culture.

\section{Related Work}

DevOps research has been largely focused on measuring performance and organizational outcomes through key indicators such as deployment frequency, lead time for changes, change failure rate, and time to restore service \cite{forsgren2018accelerate}. These so-called DORA metrics have become the industry standard for assessing software delivery performance. However, while providing high-level benchmarks, these can be misused \cite{googledorareport} and are limited in explaining the underlying causes of inefficiencies or delays. They are primarily outcome-based and lack the granularity required to understand the behavioral and procedural factors that shape these outcomes.

Process Mining \cite{vanderAalst2016} has proven to be an effective method for discovering, monitoring, and improving processes using event data extracted from information systems. Its application in the domain of software engineering is still emerging but promising. Recent studies \cite{nogueiradevops} have shown that process mining can uncover complex workflows within software development environments, revealing handovers, rework loops, and deviations from expected paths. Despite this potential, practical applications in CI/CD pipelines remain limited due to the lack of standardized event data structures and the fragmented nature of development toolchains.

In recent years, research on \emph{Predictive Process Monitoring} (PPM) has explored the use of machine learning techniques - such as gradient boosting, random forests, and deep neural networks - to predict process outcomes such as remaining time, success probability or compliance violations \cite{tax2017predictive, diFrancescomarino2024survey}. These techniques have demonstrated the ability to anticipate process behavior by learning from historical execution data. When applied to DevOps environments, similar approaches can be leveraged to forecast delivery delays or detect process anomalies \cite{predictiondevops}. Nevertheless, few works have integrated predictive analytics with real, event-level DevOps data in a production context.

Overall, current research provides a strong foundation for measuring and improving DevOps performance, but the integration of process mining and predictive modeling remains underexplored. Existing work either focuses on high-level metrics (e.g., DORA) or on isolated predictive models without contextual process information. \textbf{CodeSight} addresses this gap by unifying event data from GitHub repositories, applying process mining techniques to discover real development workflows, and augmenting these insights with machine learning models for predictive analysis.

\section{System Overview and Architecture}

\textbf{CodeSight} is an integrated system designed to predict deadline compliance in software development processes by combining data collection, process mining, analytics, and deep learning.  The system transforms raw development and deployment traces from GitHub into structured process data, enabling predictive insights over ongoing Pull Requests (PRs).

\subsection{System Components}

The architecture of CodeSight is divided into four main components:

\begin{enumerate}
    \item \textbf{Data Acquisition Layer:} retrieves development and deployment data from the GitHub REST API, including PRs, commits, and workflow runs. This component ensures traceability by linking all artifacts (branches, commits, and actions) through their SHAs.
    
    \item \textbf{Data Transformation Layer:} converts the raw data into event logs compliant with a standard process mining \textit{CSV} format. Each PR is represented as a case, and activities correspond to discrete events (e.g., PR creation, commit, merge). This enables reconstruction of actual process flows and the timing between activities.
    
    \item \textbf{Process Mining and Visualization Layer:} performs process mining, extracting elements such as workflow variants and rework patterns, as well as metrics such as lead time, review duration, or workflow success rates. These results are displayed through interactive dashboards and can be used for continuous improvement or benchmarking between repositories or teams.
    
    \item \textbf{Predictive Layer:} implements the LSTM-based model that anticipates the remaining time of a PR, using both sequential (activity traces) and static features.
\end{enumerate}

\subsection{Workflow Overview}

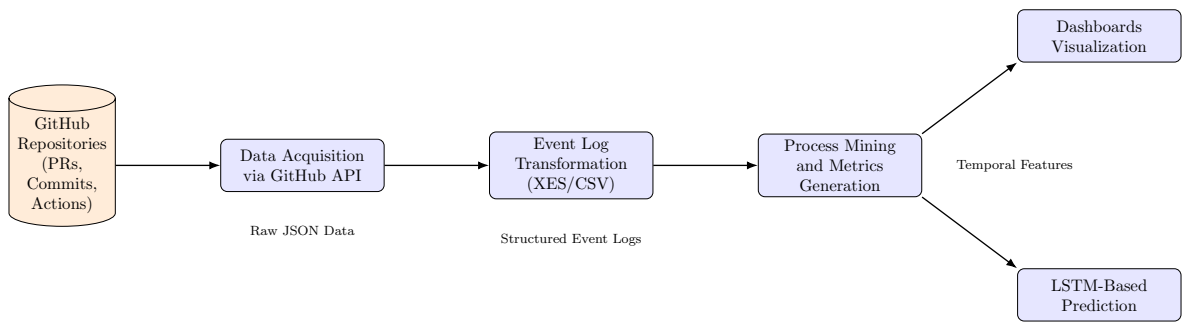
\begin{figure}[hb!]
    \centering
    \resizebox{\textwidth}{!}{%
    \begin{tikzpicture}[
        node distance=1.6cm and 2.2cm,
        process/.style={rectangle, rounded corners, draw, text width=3.2cm, minimum height=1.1cm, align=center, fill=blue!10},
        data/.style={cylinder, shape border rotate=90, draw, aspect=0.25, text width=2cm, align=center, fill=orange!15},
        arrow/.style={-Latex, thick}
    ]
    
    \node[data] (github) {GitHub Repositories \\ (PRs, Commits, Actions)};
    \node[process, right=of github] (extraction) {Data Acquisition \\ via GitHub API};
    \node[process, right=of extraction] (log) {Event Log Transformation \\ (XES/CSV)};
    \node[process, right=of log] (mining) {Process Mining \\ and Metrics Generation};
    \node[process, above right=1.5cm and 2cm of mining] (dashboards) {Dashboards \\ Visualization};
    \node[process, below right=1.5cm and 2cm of mining] (lstm) {LSTM-Based \\ Prediction};

    \draw[arrow] (github) -- (extraction);
    \draw[arrow] (extraction) -- (log);
    \draw[arrow] (log) -- (mining);
    \draw[arrow] (mining.north east) -- (dashboards.south west);
    \draw[arrow] (mining.south east) -- (lstm.north west);

    \node[below=0.6cm of extraction] {\footnotesize Raw JSON Data};
    \node[below=0.6cm of log] {\footnotesize Structured Event Logs};
    \node[right=0.6cm of mining] {\footnotesize Temporal Features};

    \end{tikzpicture}%
    }
    \caption{CodeSight architecture and data processing workflow.}
    \label{fig:architecture}
\end{figure}

The complete workflow is summarized in Figure~\ref{fig:architecture}.  
Raw GitHub data are collected and normalized, transformed into event logs, analyzed via process mining, and finally used to feed metrics dashboards and as training input for the LSTM predictor. This pipeline supports both descriptive and predictive analytics, closing the loop between process observation and actionable forecasting.

\noindent
Beyond predictive analytics, the architecture of \textbf{CodeSight} supports a continuous improvement loop for software development workflows. Predictions and process performance indicators are fed back into the organization through dashboards, alerts, and periodic reports. These insights enable development teams to identify potential deadline violations early, rebalance workloads, and refine branching or review strategies.

By combining \textit{process mining}, \textit{machine learning}, and \textit{real-time feedback}, CodeSight acts not only as a monitoring system but also as a decision-support tool that enhances both operational efficiency and software quality across development pipelines.

\section{Data Acquisition and Preparation}

The \textbf{CodeSight} system collects and transforms information directly from \textit{GitHub} repositories through its public REST API. The goal of this stage is to build a complete record of development and deployment activities that can be analyzed using process mining techniques. The following subsections describe the acquisition and transformation workflow.

\subsection{Data Extraction from GitHub}

Data are retrieved via authenticated API calls, focusing on \textit{pull requests} (PRs), their associated \textit{commits}, and \textit{workflow runs} (CI/CD executions) from a specific repository branch.

\subsubsection{Pull Requests}

The initial extraction is performed using the endpoint:

\begin{verbatim}
https://api.github.com/repos/{OWNER}/{REPO}/pulls
\end{verbatim}

All PRs are requested (\texttt{state=all}), including open, closed, and draft ones.  
The response provides core information about each PR:

\begin{itemize}
    \item \textbf{id, number, title}: unique identifiers and PR description.
    \item \textbf{login}: user who created the PR.
    \item \textbf{head\_ref}, \textbf{head\_sha}, \textbf{merge\_commit\_sha}: source branch and related commit SHAs.
    \item \textbf{base\_ref}: target branch (typically the main branch).
    \item \textbf{created\_at}, \textbf{merged\_at}, \textbf{closed\_at}: key temporal milestones.
    \item \textbf{state}, \textbf{draft}: operational status and draft flag.
    \item \textbf{assignees}, \textbf{requested\_reviewers}: developers and reviewers involved.
    \item \textbf{commits\_url}: link to detailed commit information.
\end{itemize}

Additional metadata are obtained through an individual PR call:

\begin{verbatim}
https://api.github.com/repos/{OWNER}/{REPO}/pulls/{pr_number}
\end{verbatim}

This endpoint provides complementary attributes such as  
\textbf{labels}, \textbf{merged\_by}, \textbf{commits}, \textbf{additions}, \textbf{deletions}, and \textbf{changed\_files}, 
offering a more granular view of the PR’s scope and authoring context.

\subsubsection{Commits Associated with Pull Requests}

Commit-level information is retrieved via:

\begin{verbatim}
https://api.github.com/repos/{OWNER}/{REPO}/pulls/{pr_number}/commits
\end{verbatim}

For each commit, the following attributes are collected:
\textbf{commit\_sha}, \textbf{date}, \textbf{message}, and \textbf{author}.  
Since this service does not include file details, an additional request is made for each commit:

\begin{verbatim}
https://api.github.com/repos/{OWNER}/{REPO}/commits/{commit_sha}
\end{verbatim}

This allows extraction of modified file extensions and types, enabling analysis of the technical impact of each change (e.g., code, documentation, or configuration edits).

\subsubsection{Workflow Runs (GitHub Actions)}

To capture continuous integration and deployment (CI/CD) events, workflow run data are extracted through:

\begin{verbatim}
https://api.github.com/repos/{OWNER}/{REPO}/actions/runs
\end{verbatim}

For each run, the following information is retrieved:
\textbf{id}, \textbf{name}, \textbf{head\_sha}, \textbf{event}, \textbf{status}, \textbf{conclusion}, \textbf{run\_attempt},  
\textbf{run\_started\_at}, \textbf{updated\_at}, \textbf{created\_at}, and \textbf{triggering\_actor}.  
These data link workflow executions with corresponding commits or PRs and enable calculation of metrics such as duration, success rate, and failure patterns.

\subsection{Data examples}
For our own internal tests, we extracted data from our own process mining platform development repositories in Github, namely \textit{frontend} and \textit{backend}, which were loaded into three CSV files that we describe as follows.

\subsubsection*{Dataset: exported\_commits\_inverbisanalytics\_frontend.csv}

This file contains the commit history of the frontend repository. Each row represents an individual commit performed in a branch associated with a Pull Request (PR) identified by \texttt{pr\_id}.  
The data allow analysis of source code evolution, change authorship, file types modified, and the temporal sequence of development, as Table~\ref{tab:frontend1} shows.

\begin{itemize}
    \item \textbf{Size:} 465 rows $\times$ 6 columns
    \item \textbf{Temporal coverage:} 2020-11-03 -- 2025-08-12
\end{itemize}

\begin{table}[ht!]
\centering
\caption{Commit dataset structure (frontend repository)}
\begin{tabular}{|p{3cm}|p{2.5cm}|p{8cm}|}
\hline
\textbf{Column} & \textbf{Type} & \textbf{Description} \\ \hline
\texttt{pr\_id} & ID & Unique identifier of the Pull Request associated with the commit. Groups multiple commits belonging to the same integration request. \\ \hline
\texttt{commit\_id\_sha} & text & SHA hash of the Git commit. Unique per commit, links to the full repository history. \\ \hline
\texttt{commited\_at} & datetime & Date and time when the commit was recorded in the repository. Used as the primary event timestamp. \\ \hline
\texttt{commit\_title} & text & Short commit message summarizing the purpose or main change. \\ \hline
\texttt{commit\_author} & text & User or author who performed the commit. May contain null values for automated or missing commits. \\ \hline
\texttt{filetypes} & text & List or set of modified file types (e.g., \texttt{.js}, \texttt{.vue}, \texttt{.css}, \texttt{.yaml}). \\ \hline
\end{tabular}
\label{tab:frontend1}
\end{table}

\textbf{Subsequent use:} this dataset provides the “Commit” events in the process mining trace, linked to the same \texttt{pr\_id} as PR opening or merging events.

\subsubsection*{Dataset: exported\_inv\_frontend\_some\_missing.csv}

This dataset contains Pull Request (PR) and workflow execution information from the frontend repository. Each row corresponds to a combination between a PR (\texttt{pr\_id}) and a possible associated pipeline execution (\texttt{run\_id}).  
It includes authorship, state, key timestamps (creation, merge, closure), and change metrics. Table~\ref{tab:frontend2} lists all the data fields we extract.

\begin{itemize}
    \item \textbf{Size:} 1179 rows $\times$ 32 columns
    \item \textbf{Temporal coverage:} 2023-02-13 -- 2025-08-27
\end{itemize}

\begin{table}[ht!]
\centering
\caption{PR and workflow dataset structure (frontend repository)}
\begin{tabular}{|p{3cm}|p{2.5cm}|p{8cm}|}
\hline
\textbf{Column} & \textbf{Type} & \textbf{Description} \\ \hline
\texttt{pr\_id} & ID & Internal identifier of the Pull Request (primary case key). \\ \hline
\texttt{pr\_number} & numeric & PR number assigned by GitHub. \\ \hline
\texttt{pr\_title} & text & Descriptive title summarizing the purpose of the PR. \\ \hline
\texttt{pr\_author} & text & User who created the PR. \\ \hline
\texttt{from\_branch} & text & Source branch from which the merge was proposed (author’s branch). \\ \hline
\texttt{head\_sha} & text & SHA hash of the last commit before merging. \\ \hline
\texttt{merge\_commit\_sha} & text & SHA of the merge commit, when the PR was successfully merged. \\ \hline
\texttt{into\_branch} & text & Target branch where the PR is merged. \\ \hline
\texttt{created\_at\_x} & datetime & PR creation timestamp. \\ \hline
\texttt{merged\_at} & datetime & Merge timestamp; null if closed without merge. \\ \hline
\texttt{closed\_at} & datetime & PR closure date (either merged or cancelled). \\ \hline
\texttt{state} & text & Final state of the PR (\texttt{open}, \texttt{closed}, \texttt{merged}). \\ \hline
\texttt{is\_draft} & boolean & Indicates whether the PR was marked as draft. \\ \hline
\texttt{assignees} & text & List of users assigned to the PR. \\ \hline
\texttt{reviewers} & text & List of requested reviewers. \\ \hline
\texttt{merged\_by} & text & User who performed the merge action. \\ \hline
\texttt{commits} & integer & Total number of commits included in the PR. \\ \hline
\texttt{additions} & integer & Number of added lines. \\ \hline
\texttt{deletions} & integer & Number of deleted lines. \\ \hline
\texttt{changed\_files} & integer & Number of modified files. \\ \hline
\texttt{labels} & text & Labels or tags applied to the PR. \\ \hline
\texttt{run\_id} & ID & Identifier of the pipeline execution. \\ \hline
\texttt{run\_name} & text & Workflow name. \\ \hline
\texttt{run\_pr\_commit \_head\_sha} & text & SHA of the commit associated with the workflow run. \\ \hline
\texttt{event\_trigger} & text & Event type that triggered the pipeline execution. \\ \hline
\texttt{status} & text & Execution status. \\ \hline
\texttt{conclusion} & text & Final result (success, failure, etc.). \\ \hline
\texttt{created\_at\_y} & datetime & Creation timestamp of the run record. \\ \hline
\texttt{run\_attempt} & integer & Attempt number of the execution. \\ \hline
\texttt{run\_started\_at} & datetime & Start time of the pipeline run. \\ \hline
\texttt{duration\_ms} & numeric & Total pipeline duration in milliseconds. \\ \hline
\texttt{actor\_trigger} & text & User or process that triggered the pipeline. \\ \hline
\end{tabular}
\label{tab:frontend2}
\end{table}

\textbf{Subsequent use:} this dataset provides the “PR Opening”, “PR Merge”, “PR Closure”, and workflow-related events.

\subsubsection*{Dataset: exported\_inv\_backend.csv}

This dataset has the same structure as the previous one but corresponds to the backend repository. It also includes individual commits associated with PRs.

\begin{itemize}
    \item \textbf{Size:} 1375 rows $\times$ 37 columns
    \item \textbf{Temporal coverage:} 2024-07-29 -- 2025-06-04
\end{itemize}

\textbf{Subsequent use:} this dataset enables tracing of PR opening, closure, and merge events, as well as workflow executions and individual commits.

\subsection{Data Transformation and Event Log Generation}

The transformation process aims to structure the raw data exported from GitHub into a structured event log compatible with process mining tools.

This conversion is performed using a set of Python scripts organized into three functional blocks:

\paragraph{Trace Structure Generation.}  
Date columns are converted to \texttt{datetime} format, and relevant attributes are selected for each type of table (commitments, PRs, and workflows). Then, the tabular data are transformed into a sequential event model, where each case (\texttt{pr\_id}) can have multiple activities: PR opening, commits, workflow executions, merge, or closure.  
The procedure identifies all columns with the prefix \texttt{"Fch"} (event dates), groups records by \texttt{pr\_id}, and for each detected date column creates a new row duplicating the PR metadata.  

For each new row:
\begin{itemize}
    \item The date value is stored in the \texttt{DATE} column.
    \item The original column name is assigned to \texttt{ACTIVITY}, which is later translated into a readable activity name.
\end{itemize}

Finally, the resulting table is sorted chronologically by \texttt{pr\_id} and \texttt{DATE} and reindexed. Each row thus represents an activity with its timestamp linked to the corresponding case.

\paragraph{Attribute Configuration and Selection}  
In this phase, the columns for the final dataset are defined (identifier, date, activity type, and relevant attributes).  
Activity names extracted from the date columns are translated into more understandable terms (e.g., \texttt{Fch commit} $\rightarrow$ \texttt{Commit}, \texttt{Fch apertura PR} $\rightarrow$ \texttt{PR Opening}).

\paragraph{Execution and Integration of the Data Model}  
The three original CSV files are loaded and the date columns are renamed with descriptive labels. The transformations described previously are applied and the resulting traces are merged into a single unified data set containing the following main fields:
\begin{itemize}
    \item \texttt{pr\_id} (case identifier)
    \item \texttt{ACTIVITY} (event)
    \item \texttt{DATE} (timestamp)
    \item User attributes: \texttt{commit\_author}, \texttt{pr\_author}, \texttt{merged\_by}, etc.
    \item Additional context attributes: \texttt{from\_branch}, \texttt{filetypes}, \texttt{state}, etc.
\end{itemize}

\begin{figure}
\centering
\includegraphics[width=0.8\linewidth]{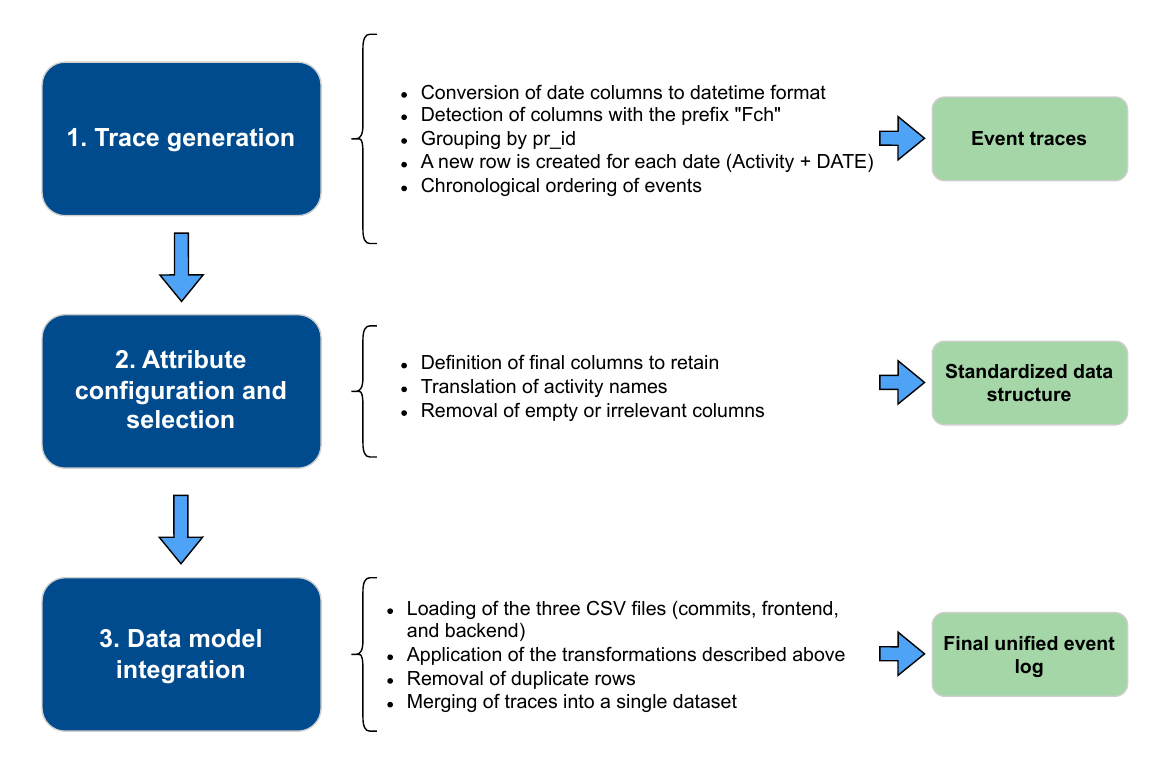}
\caption{\label{fig:transformation} Conversion of raw data}
\end{figure}

This process, summarized by Figure~\ref{fig:transformation}, standardizes heterogeneous structures into a coherent single model, translates and contextualizes activities for interpretation, and produces an event log directly exportable to process mining tools.

\subsection{Resulting Dataset}  
After applying the transformations, a unified event log dataset is obtained, where each row represents an activity performed during a Pull Request (PR) lifecycle.  
The identifier \texttt{pr\_id} acts as the case ID, grouping all activities associated with the same PR from opening to closure.

\section{Process Mining and Dashboards}
This section details the use of process mining and dashboarding techniques to analyze pull request event data. The aim is to reconstruct workflow structures, identify bottlenecks, and visualize key performance indicators through a Power BI dashboard. These insights serve both process optimization and as input for the predictive modeling presented in Section 6.

\subsection{Mining the event log}
The resulting event log is analyzed using our own \textbf{process mining} platform to reconstruct workflow structures and compute operational metrics, including:

\begin{itemize}
    \item Activity and transition durations (e.g., time from PR creation to merge).
    \item Waiting and review times.
    \item Identification of bottlenecks and rework patterns.
    \item Performance indicators such as throughput and deadline compliance.
\end{itemize}

The process we obtained from our own data comprises a total of 835 cases and 271 process variants, with a temporal range from 2020-11-03 to 2025-08-27. The high number of different paths indicates substantial variability in process execution, which is typical in software development projects with multiple branches, authors, and automation workflows. The average process duration is 7 days and 14 hours, highlighting notable differences between cases depending on complexity or number of iterations.

\begin{figure}[ht]
\centering
\includegraphics[width=0.75\linewidth]{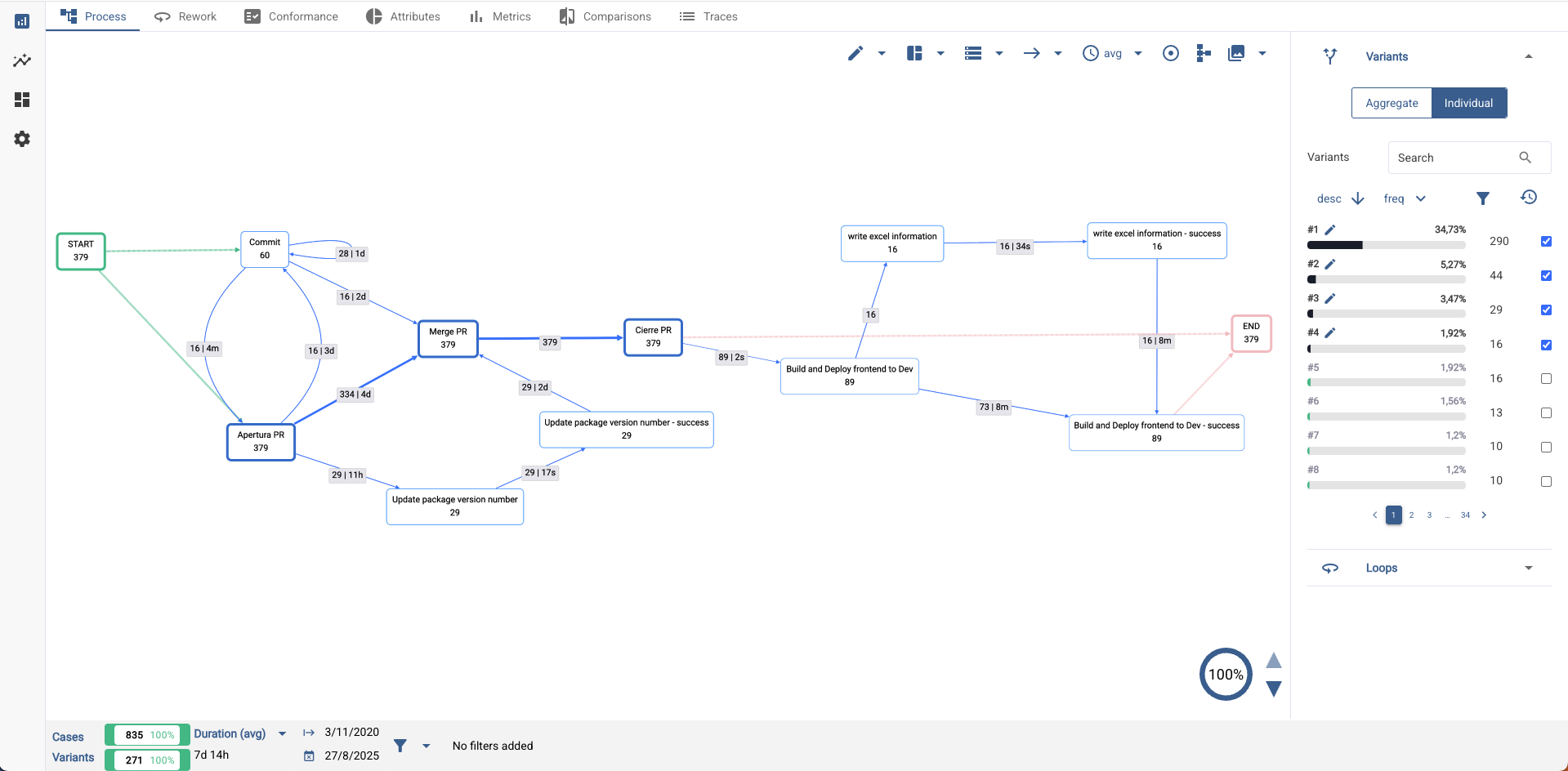}
\caption{\label{fig:pmining}Simplified process model.}
\end{figure}

This process model, shown in Figure~\ref{fig:pmining}, enables identification of improvement patterns, analysis of dependencies between activities, and detection of potential bottlenecks in code review or automation phases. Likewise, the temporally structured traces enriched with derived metrics serve as the direct input for both the insights dashboard and the LSTM-based predictive model described in the next section.

\subsection{Insights Dashboard}

The CodeSight system includes a comprehensive Power BI dashboard designed to monitor software development and deployment processes through key performance indicators and process mining results. This dashboard feeds directly from the Inverbis Analytics process mining platform's API.
The dashboard is organized into seven main sections, each providing a distinct analytical perspective:

\begin{enumerate}
    \item \textbf{DORA Metrics:}  
    This first tab presents the core DORA metrics (Deployment Frequency, Lead Time for Changes, Change Failure Rate, and Mean Time to Restore). These indicators provide a high-level view of the software delivery performance and reliability. These are shown in Figure~\ref{fig:dashboard1}.

    \begin{figure}[ht]
    \centering
    \includegraphics[width=0.75\linewidth]{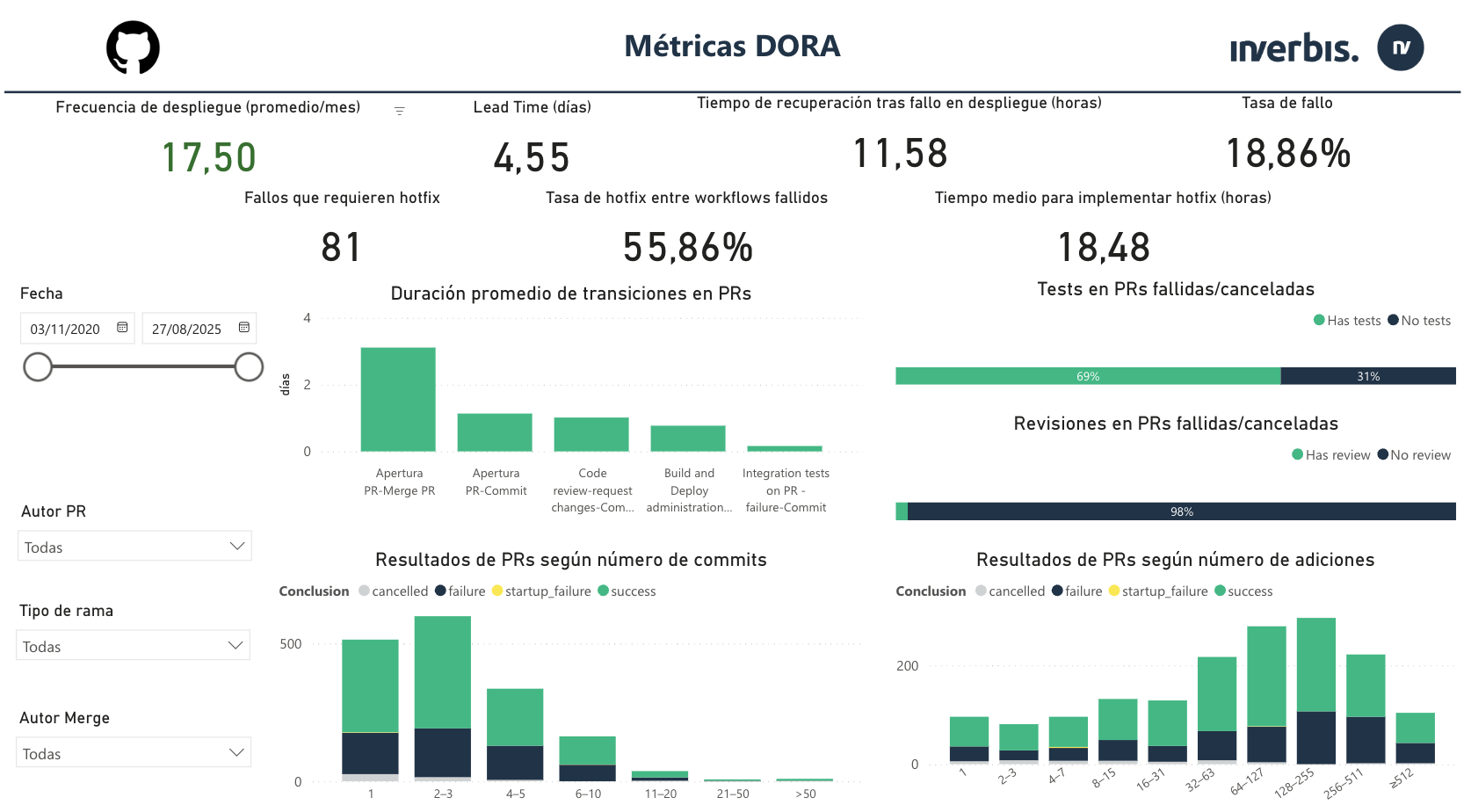}
    \caption{\label{fig:dashboard1} DORA Metrics view }
    \end{figure}

    \item \textbf{General Development Indicators:}  
    The second tab, shown in Figure~\ref{fig:dashboard2}, provides general insights into development activity, including deployment frequency, lead time, average implementation time, and post-merge duration. It also includes distributions of merge events by author, offering visibility into developer participation and workload.

    \item \textbf{Pull Request Activity:}  
    This section focuses on Pull Request (PR) dynamics, presenting the number of PRs created, average review times, and distributions by author and duration. Additional visualizations highlight the average PR lifetime per author, supporting analysis of collaboration and code review efficiency.

    \begin{figure}[ht]
    \centering
    \includegraphics[width=0.75\linewidth]{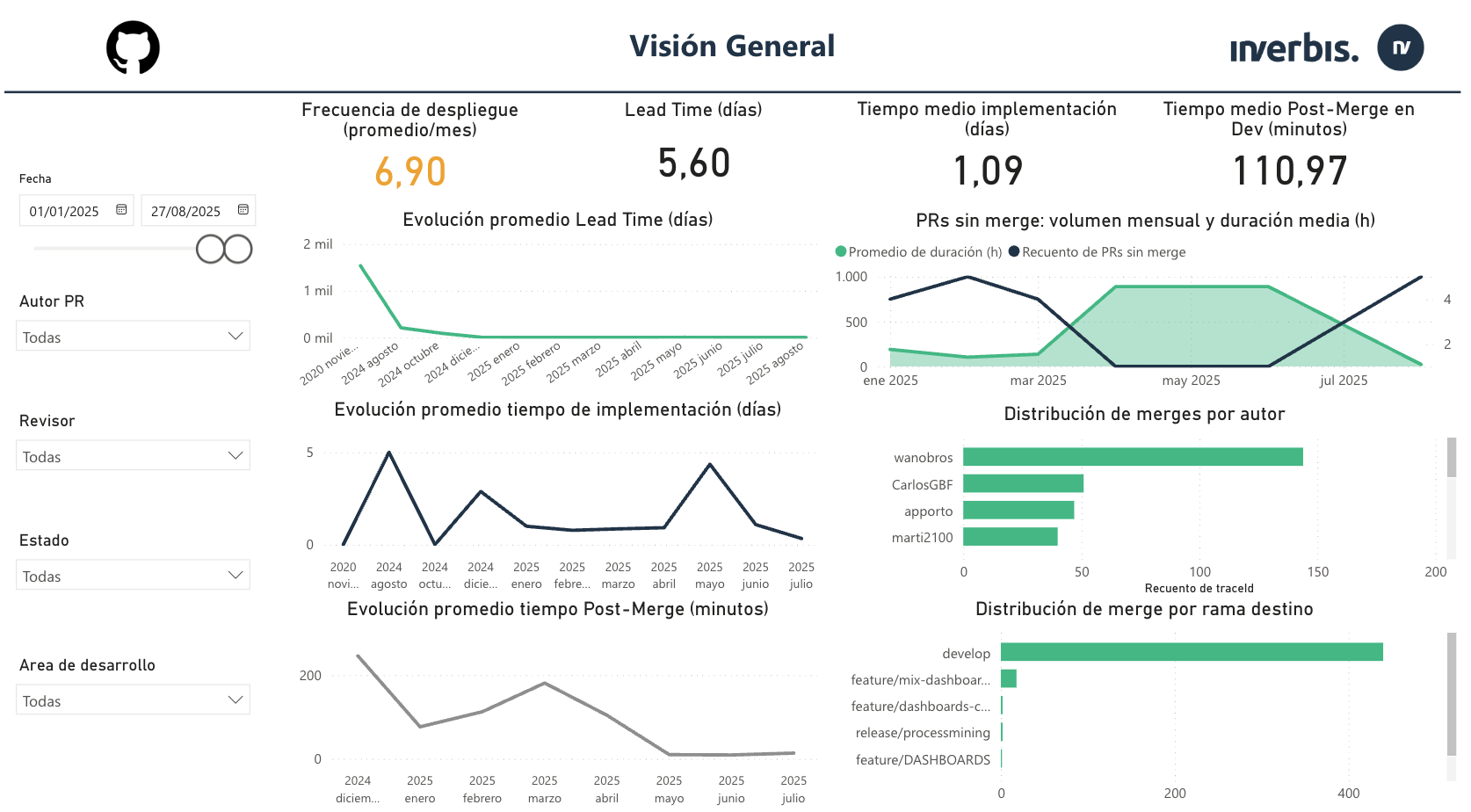}
    \caption{\label{fig:dashboard2} General development indicators view.}
    \end{figure}

    \item \textbf{Process Variants and Visualization:}  
    The fourth tab integrates process mining outputs, displaying the discovered process variants and an interactive visualization of the PR lifecycle. This allows users to identify deviations, redundant paths, or bottlenecks in the workflow.

    \item \textbf{User-based Analysis:}  
    This view provides a per-user analysis of activity across PRs, commits, and workflows, enabling identification of development patterns, workload balance, and team-level performance.

    \item \textbf{Temporal Evolution of PRs:}  
    The sixth tab explores the temporal trends of PR creation and integration, showing the evolution of activity over time along with complementary indicators such as average completion time and throughput.

    \item \textbf{Deployment and Incident Metrics:}  
    The final section focuses on deployment-related information, including failed workflow counts, incident rates, breakdowns by author and repository type, and the monthly evolution of incidents. This view connects development dynamics with operational outcomes, closing the DevOps feedback loop.
\end{enumerate}

Together, these dashboards provide an integrated analytical environment for continuous monitoring of software development processes, combining process mining insights with operational metrics for decision support and process improvement.

\section{LSTM for Deadline Compliance Prediction in Pull Requests}
This section describes the methodology used to develop and evaluate a predictive model for deadline compliance in pull requests (PRs). The goal is to determine whether a PR will meet its assigned deadline based on historical process data and contextual attributes.

To achieve this, a Long Short-Term Memory (LSTM) neural network is implemented, as it can effectively capture the temporal dependencies within sequences of PR activities. We adopted the use of LSTM networks since these are well-suited for modeling temporal dependencies in sequential data, such as the ordered activities within a pull request lifecycle. The section covers all stages of the modeling pipeline, from data preparation to model evaluation.

First, the feature generation and preprocessing steps are described, including the creation of categorical, binary, and numerical variables, as well as methods to prevent data leakage and transform target variables. Then, the training pipeline is presented, encompassing dataset splitting, sequence padding, and temporal normalization. Finally, the LSTM model architecture, training configuration, and evaluation metrics are detailed, followed by the presentation and discussion of the results.

\subsection{Feature Generation and Selection}

\subsubsection*{Branch Attributes}
\begin{itemize}
    \item \texttt{from\_branch\_type}: type of the PR’s source branch (fix, hotfix, bug, feature).
    \item \texttt{into\_branch\_type}: type of the target branch (feature, develop, release, staging).
    \item \texttt{process}: distinguishes between backend/frontend.
\end{itemize}

\subsubsection*{File Types}
A binary column \texttt{has\_X} is generated for each file type found in the original \texttt{filetypes} column, with a value of 1 if that type is modified and 0 otherwise.

\subsubsection*{Activity Encoding}
Each activity name is converted into an integer (reserving 0 for padding). For each PR, a sequential list of indices of the executed activities is obtained.

\subsubsection*{Temporal Attributes}
From the \texttt{DATE} column, the following features are created: \texttt{year}, \texttt{month}, \texttt{day}, \texttt{weekday}, and \texttt{is\_weekend}.

\subsubsection*{Durations Between Activities}
\texttt{transition\_seconds} is calculated, a list of elapsed times between consecutive activities.

\subsubsection*{Labels and Deadline}
Each PR is classified by complexity level: S, M, L, or XL. Each level is assigned a deadline in hours:
\begin{itemize}
    \item S: 8 h
    \item M: 24 h
    \item L: 48 h
    \item XL: 72 h
\end{itemize}

\subsection{Trace Truncation}
To simulate the “in-progress” state of a PR and predict its evolution, each trace is truncated at a random position, generating new attributes:
\begin{itemize}
    \item \texttt{cut}: index at which the trace is truncated.
    \item \texttt{truncated\_activity\_list}: activities up to the truncation point.
    \item \texttt{remaining\_activity\_list}: remaining activities.
    \item \texttt{truncated\_transitions} and \texttt{remaining\_transitions}: durations corresponding to each segment.
    \item \texttt{elapsed\_time}: total time elapsed up to the truncation.
    \item \texttt{remaining\_time}: remaining time until the actual closure.
    \item \texttt{activity}: index of the last activity recorded before truncation.
\end{itemize}

Other metrics derived from the prefix are also calculated:
\begin{itemize}
    \item \texttt{prefix\_len}: length of the prefix.
    \item \texttt{trunc\_dt\_mean}: mean time between events in the prefix.
\end{itemize}

\subsection{Data Preparation and Feature Engineering}
\subsubsection*{Preparation of Variables for Training}
\begin{itemize}
    \item \textbf{Prevention of data leakage:} Variables that reveal future information or are not constant along the trace (e.g., \texttt{trace\_total\_duration} or \texttt{remaining\_transitions}) are removed.
    \item \textbf{Target variable transformation:} A logarithmic transformation (\texttt{log1p}) is applied to \texttt{remaining\_time} to reduce skewness and stabilize training.
\end{itemize}

\subsubsection*{Feature Selection}
\begin{itemize}
    \item \textbf{Continuous numerical:} \texttt{elapsed\_time}, \texttt{prefix\_len}, and \texttt{trunc\_dt\_mean}.
    \item \textbf{Categorical:} calendar variables, branch type, and activity type.
    \item \textbf{Binary:} \texttt{has\_} columns associated with events occurring before the cut.
\end{itemize}

A complete list of features (\texttt{all\_features}) is compiled, and categorical variables are typed as strings for proper encoding.

\subsubsection*{Column Preprocessing}
\begin{itemize}
    \item \textbf{Numerical:} missing values are imputed with the median and scaled using \texttt{StandardScaler}.
    \item \textbf{Binary:} left unchanged.
    \item \textbf{Categorical:} encoded using \texttt{OneHotEncoder}.
\end{itemize}

These steps are integrated into a \texttt{ColumnTransformer}, which is later included in a general pipeline (\texttt{prep\_pipeline}).

\subsection{Dataset Splitting}
The dataset is split approximately into:
\begin{itemize}
    \item 70\% training
    \item 15\% validation
    \item 15\% test
\end{itemize}

Explanatory variables (\texttt{X\_train}, \texttt{X\_val}, \texttt{X\_test}) and transformed target labels (\texttt{y\_train\_log}, \texttt{y\_val\_log}, \texttt{y\_test\_log}), along with their original versions in seconds (\texttt{y\_train}, \texttt{y\_val}, \texttt{y\_test}), are extracted.

The pipeline is fitted on the training set (\texttt{fit\_transform}) and then applied to validation and test sets (\texttt{transform}) without retraining, ensuring consistency and avoiding data leakage.

\subsection{Sequence Padding}
To ensure uniform sequence lengths, the end of the activity list is padded with 0. The maximum length is set at the 95th percentile to prevent excessively long sequences from unnecessarily increasing the tensor size. For the transition list, a 0 is added at the beginning of each list (representing time before the first event).

\subsection{Duration Transformation}
Durations between activities are positive values that can vary greatly. They are transformed using a logarithm and standardized. The transformation also returns parameters (\texttt{mu}, \texttt{sd}) to apply later to the validation and test sets.

\subsection{Model architecture}
\begin{figure}[ht!]
\centering
\includegraphics[width=0.75\linewidth]{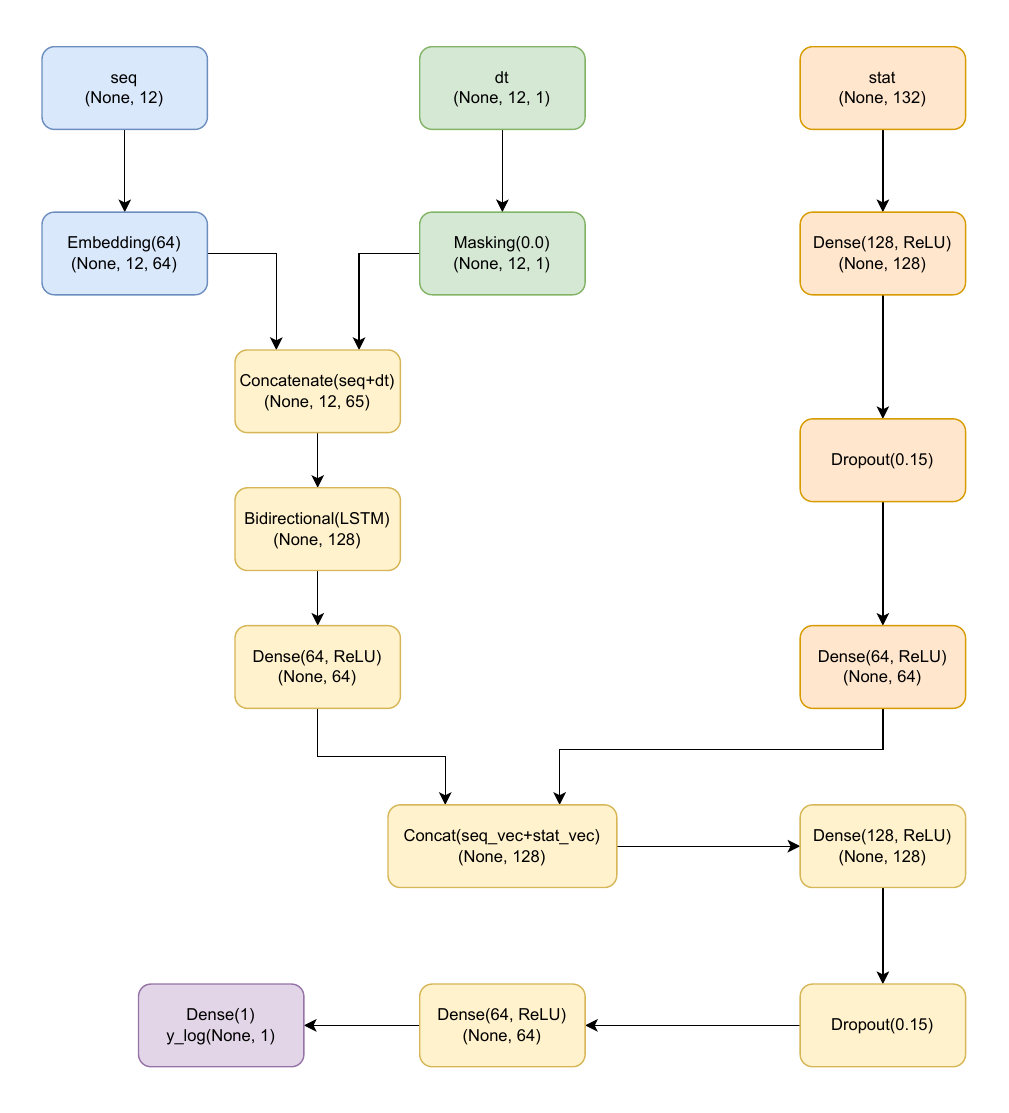}
\caption{\label{fig:lstmarchitecture} Architecture of the LSTM model}
\end{figure}

The architecture of the LSTM model is shown in Figure \ref{fig:lstmarchitecture}.
\subsubsection*{Model Inputs}
\begin{itemize}
    \item \texttt{seq\_in}: sequence of activity IDs
    \item \texttt{dt\_in}: transition times associated with those activities, of the same length
    \item \texttt{stat\_in}: processed static features (numerical and categorical)
\end{itemize}

\subsubsection*{Sequential Branch: Activities + Transition Times}
\begin{itemize}
    \item \textbf{Embedding layer} (64 dimensions, ignores padding)
    \item \textbf{Transition mask}: ignores zeros
    \item \textbf{Concatenation} of embedding + duration
    \item \textbf{BiLSTM}: reads the full sequence (forward and backward) and learns temporal patterns. Each direction has 64 neurons, combined into a final vector of 128 values
    \item \textbf{Dropout = 0.15} randomly disables some connections during training to prevent overfitting
\end{itemize}

\subsubsection*{Sequential Information Compression}
\begin{itemize}
    \item Dense layer with 64 neurons and ReLU activation
\end{itemize}

\subsubsection*{Static Feature Branch}
\begin{itemize}
    \item Small dense network processing non-sequential information (numerical and categorical)
    \item Dense(128) - Dropout(0.15) - Dense(64)
\end{itemize}

\subsubsection*{Fusion of Both Branches}
\begin{itemize}
    \item The features are merged into a layer of 128 neurons.
    \item A more compact dense layer with 64 neurons is applied.
    \item A Dropout layer with a 15\% rate is applied.
\end{itemize}

\subsubsection*{Prediction Head}
\begin{itemize}
    \item Combines all learned signals to predict the final value
    \item Outputs a continuous value representing the logarithm of the remaining time
\end{itemize}

\subsection{Compilation and Training}
\begin{itemize}
    \item \textbf{Optimizer}: Adam with learning rate $1\mathrm{e}{-3}$
    \item \textbf{Loss function}: mean squared error (MSE) on log-transformed values, penalizing proportionally large relative errors in long durations; mean absolute error (MAE) is also added for an intuitive reference during training.
    \item \textbf{Callbacks}:
    \begin{itemize}
        \item \texttt{EarlyStopping}: stops training if validation loss does not improve for 8 consecutive epochs, restoring best weights.
        \item \texttt{ReduceLROnPlateau}: halves the learning rate when validation loss stagnates for 3 epochs, with a minimum of $1\mathrm{e}{-5}$, helping fine-tune the final stages of learning.
    \end{itemize}
    \item The model is trained using all three input types (activity sequence, transition sequence, and static features).
    \item Maximum of 40 epochs, batch size of 64, validating on the validation set and applying defined callbacks to control convergence and overfitting.
\end{itemize}

\subsection{Model Performance Evaluation}
\begin{itemize}
    \item Predictions are generated and the logarithmic transformation is reversed to obtain durations in seconds.
    \item MAE is calculated between predictions and actual values, converting to hours.
    \item Metrics are evaluated on the log scale; the coefficient of determination ($R^2$) is calculated to indicate the percentage of explained variance.
    \item Deadline compliance is evaluated by checking whether total elapsed time (\texttt{elapsed\_time + predicted\_duration}) stays within the maximum allowed limit (\texttt{Deadline\_hours * 3600}).
    \item Compared with actual outcomes to obtain binary metrics: accuracy and F1-score, indicating the model's ability to correctly anticipate whether a case will meet the deadline.
\end{itemize}

\subsection{Results}

\subsubsection*{Regression Metrics}
The target variable was trained on a logarithmic scale. The following metrics are reported in Table~\ref{tab:regression}:

\begin{itemize}
    \item \textbf{MAE(h)}: mean absolute error in hours (original scale)
    \item \textbf{R2(log)}: coefficient of determination on the log scale
\end{itemize}

\begin{table}[ht]
\centering
\caption{\label{tab:regression}Regression performance metrics.}
\begin{tabular}{lcc}
\toprule
Set & MAE(h) & R2(log) \\
\midrule
Train & 10.108 & 0.938 \\
Validation & 11.627 & 0.843 \\
Test & 8.801 & 0.781 \\
\bottomrule
\end{tabular}
\end{table}

The model explains between 80\% and 93\% of the variance. The MAE of 8.8 hours on the test set represents roughly one working day of deviation, which is acceptable for practical deadline forecasting purposes.

\subsubsection*{Deadline Compliance Evaluation}
Predicted remaining durations are transformed to seconds and compared with the deadline limit (\texttt{deadline\_hours} $\times$ 3600). Accuracy and F1-score are reported in Table~\ref{tab:scores}.  

A case is predicted as deadline-compliant if:
\[
\texttt{elapsed\_time} + \texttt{predicted\_duration} \leq \texttt{deadline\_hours} \times 3600
\]

\begin{table}[h]
\centering
\caption{\label{tab:scores}Deadline compliance classification results (Accuracy and F1-score)}
\begin{tabular}{lcc}
\toprule
Set & Accuracy & F1 \\
\midrule
Train & 0.940 & 0.959 \\
Validation & 0.881 & 0.911 \\
Test & 0.944 & 0.963 \\
\bottomrule
\end{tabular}

\end{table}

On the test set, the model achieves an accuracy of 0.944 and F1-score of 0.963, indicating a high ability to anticipate deadline compliance from incomplete traces.

\begin{figure}[ht!]
\centering
\includegraphics[width=0.6\linewidth]{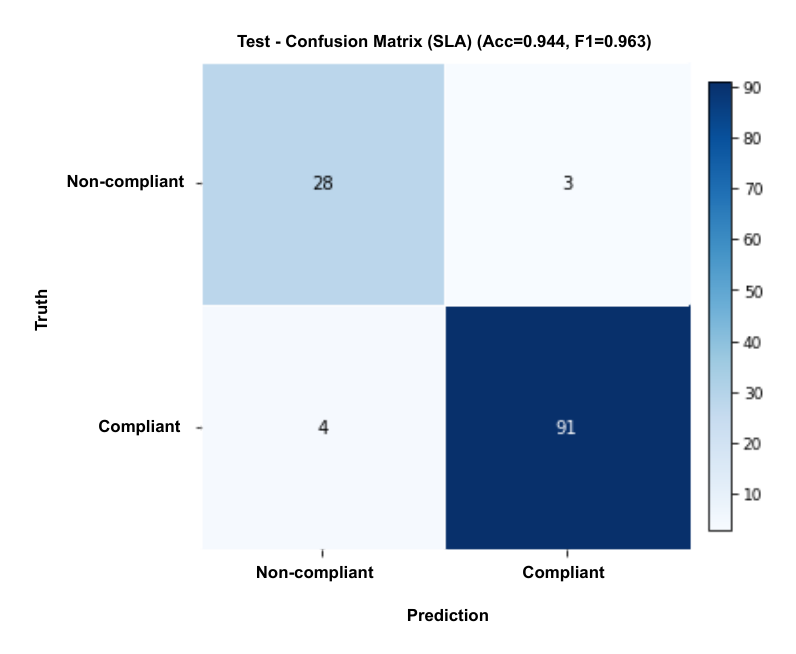}
\caption{\label{fig:confusion}Test confusion matrix}
\end{figure}

The confusion matrix on the test set (Figure \ref{fig:confusion} shows very strong performance: 91 true positives (TP), 28 true negatives (TN), 3 false positives (FP), and 4 false negatives (FN).  

\begin{itemize}
    \item Predicted “Compliant” precision: $\approx 91/(91+3) = 0.968$
    \item Sensitivity (recall) for “Compliant”: $\approx 91/(91+4) = 0.958$
    \item Specificity for “Non-compliant”: $\approx 28/(28+3) = 0.903$
\end{itemize}

Operationally, only 3 cases were falsely authorized as “Compliant” (low risk), and 4 were marked as “Non-compliant” despite actually meeting the deadline (slightly conservative criterion).

The consistency between the continuous remaining-time predictions and the binary deadline compliance outcomes indicates that the model successfully captures the underlying temporal dynamics of the process. This coherence supports the methodological decision to employ a unified continuous output for both regression and classification perspectives.

\section{Conclusions and Future Work}
This work presents \textit{CodeSight}, a system that integrates data acquisition, process mining, metric generation, and predictive modeling to analyze software development workflows. The system demonstrates how process-oriented representations of software development activities---derived from GitHub repositories---can be leveraged to build interpretable metrics and predictive models that anticipate project outcomes.

The results obtained confirm that the approach is capable of modeling complex temporal and structural patterns within development processes, achieving high predictive performance when estimating task completion times and deadline compliance. Moreover, the combination of process mining with LSTM-based predictive models provides a comprehensive view that supports both operational monitoring and forward-looking insights.

However, several challenges remain open for future work. First, companies rely on different code repository services (e.g., GitLab, Bitbucket, Azure DevOps), each with its own API structure and data semantics. Extending \textit{CodeSight} to integrate multiple platforms would enhance its generalization and applicability across heterogeneous environments. Second, even within the same platform, development teams often organize their workflows differently---using distinct branching strategies, labels, or issue linking conventions. Therefore, designing a flexible data extraction layer capable of adapting to diverse repository structures will be essential to ensure the robustness and portability of the approach.

Future work will focus on three main directions: 
(i) expanding data integration to additional repository services and CI/CD systems; 
(ii) automating the mapping of repository events into standardized process logs; and 
(iii) incorporating explainability mechanisms into predictive models to facilitate their adoption in industrial environments.
(iv) splitting the prediction approach into development and deployment for more specific estimations of both stages.

Ultimately, \textit{CodeSight} represents a step toward process intelligence for DevOps, enabling organizations to continuously monitor, understand, and predict the behavior of their development pipelines through data-driven insights.

\section*{Acknowledgements}
This work has been funded by the Galician Institute for Economical Promotion (IGAPE) through grant IG408M-2025-000-000206, by the Spanish Ministry of Economic Matters and Digital Transformation and the European Union (Next Generation funds).

\bibliographystyle{alpha}
\bibliography{bibliography}

\end{document}